\documentstyle[11pt]{article}
\begin{document}
{\setlength{\oddsidemargin}{1.5in}
\setlength{\evensidemargin}{1.5in} } \baselineskip 0.50cm
\begin{center}
{\LARGE {\bf Non-stationary de Sitter cosmological models}}
\end{center}
\begin{center}Ng. Ibohal \\
Department of Mathematics, Manipur University,\\
Imphal 795003, Manipur, INDIA.\\
E-mail: ngibohal@iucaa.ernet.in
\end{center}
\date{}

\begin{abstract}
In this note it is proposed a class of non-stationary de Sitter,
rotating and non-rotating, solutions of Einstein's field equations
with a cosmological term of variable function $\Lambda^*(u)$. It
is found that the space-time of the rotating non-stationary de
Sitter model is an algebraically special in the Petrov
classification of gravitational field with a null vector, which is
geodesic, shear free, expanding as well as non-zero twist.
However, that of the non-rotating non-stationary model is
conformally flat with non-empty space. \\\
\end{abstract}
\vspace*{8.8pt}

It is well known that the original de Sitter cosmological model is
{\sl conformally} flat $C_{abcd}=0$ space-time with {\sl constant
curvature} $R_{abcd}=(\Lambda^*/3)(g_{ac}g_{bd}-g_{ad}g_{bc})$
[1]. It also describes the {\sl non-rotating} and {\sl stationary}
solution. Therefore, the non-rotating stationary de Sitter model
is a solution of Einstein's field equations for an empty space
with constant curvature, whereas the {\sl rotating} stationary de
Sitter model proposed in Ref. 2 is a solution for {\sl non-empty}
space with {\sl non-constant} curvature. Because of the stationary
and non-rotating properties of the original de Sitter space, the
non-rotating Schwarzschild black hole with constant mass can embed
to produce Schwarzschild-de Sitter cosmological black hole with
two event horizons - one for black hole and other for cosmological
[3]. Similarly, the rotating stationary de Sitter cosmological
universe [2] can conveniently embed into the rotating stationary
Kerr-Newman solution to produce rotating Kerr-Newman-de Sitter
cosmological black hole with constant cosmological term. This
Kerr-Newman-de Sitter black hole metric can be expressed in terms
of Kerr-Schild ansatz with different backgrounds as $g_{ab}^{\rm
KNdS}=g_{ab}^{\rm dS} +2Q(r,\theta)\ell_a\ell_b$ where
$Q(r,\theta) =-(rm-e^2/2)R^{-2}$, and $g_{ab}^{\rm
KNdS}=g_{ab}^{\rm KN}+2H(r,\theta)\,\ell_a\,\ell_b$  with
$H(r,\theta)=-(\Lambda^*r^4/6)\,R^{-2}$. Here $g_{ab}^{\rm dS}$ is
the rotating stationary de Sitter metric and the vector $\ell_a$
is a geodesic, shear free, expanding as well as non-zero twist,
and one of the repeated principal null vectors of $g^{\rm
KN}_{ab}$, $g_{ab}^{\rm dS}$ and $g_{ab}^{\rm KNdS}$, as these
space-times are Petrov type D. The expressibility of an embedded
black hole in different Kerr-Schild ansatze means that, it is
always true to talk about either Kerr-Newman black hole embedded
into the rotating de Sitter space as Kerr-Newman-de Sitter or the
rotating de Sitter space into Kerr-Newman black hole as rotating
de Sitter-Kerr-Newman black hole - geometrically both are the
same. That is, physically one may not be able to predict which
space starts first to embed into what space. One thing we found
from the study of Hawking's radiation of Kerr-Newman-de Sitter
black hole [4], is that, there is no effect on the cosmological
constant $\Lambda^*$ during the evaporation process of electrical
radiation. The cosmological constant $\Lambda^*$ always remains
unaffected in Einstein's field equations during Hawking's
radiation process. That is, unless some external forces apply to
remove the cosmological term $\Lambda^*$ from the space-time
geometry, it continues to exist along with the electrically
radiating objects, rotating or non-rotating. This means that it
might have started to embed from the very early stage of the
embedded black hole, and should continue to embed forever. It is
noted that the Kerr-Newman-de Sitter black hole proposed in Ref. 2
is found different from the one obtained by Carter [5] in the
terms involving cosmological constant.

The black hole embedded into de Sitter space plays an important
role in classical general relativity that the cosmological
constant is found present in the inflationary scenario of the
early universe in a stage where the universe is geometrically
similar to the original de Sitter space [6]. Also embedded black
holes can avoid the direct formation of negative mass naked
singularities during Hawking's black hole evaporation process [4].
It is also known that the rotating Vaidya-Bonnor black hole with
variable mass $M(u)$ and charge $e(u)$ is a non-stationary
solution. When $M(u)$ and $e(u)$ become constants, the rotating
Vaidya-Bonnor black hole will reduce to the stationary Kerr-Newman
black hole. If one wishes to study the physical properties of the
gravitational field of a {\sl complete non-stationary} embedded
black hole, e.g. rotating non-stationary Vaidya-Bonnor-de Sitter
(not discussed in this note), one needs to derive a new {\sl
rotating non-stationary} de Sitter model with a cosmological term
of variable function $\Lambda^*(u)$. That is, an observer
traveling in a non-stationary space-time must also be able to find
a non-stationary cosmological de Sitter space to embed, having a
similar space-time structure with time dependent functions.

In this view, it is proposed a rotating {\sl non-stationary} de
Sitter solution of Einstein's field equations with a cosmological
term of variable function $\Lambda^*(u)$ in this note. Using
Newman-Penrose formalism [7], a class of rotating metric with a
mass function $M(u,r)$ has been discussed in Ref. 2, where the
mass function is being expressed in terms of Wang-Wu function
$q_n(u)$ [8] as
\begin{eqnarray*}
M(u,r)\equiv \sum_{n=-\infty}^{+\infty} q_{\,n}(u)\,r^n.
\end{eqnarray*}
For obtaining a rotating {\sl non-stationary} de Sitter solution,
we choose the Wang-Wu function as
\begin{eqnarray}
\begin{array}{cc}
q_n(u)=&\left\{\begin{array}{ll}
\Lambda^*(u)/6, &{\rm when}\;\;n=3\\
0, &{\rm when }\;\;n\neq 3,
\end{array}\right.
\end{array}
\end{eqnarray}
such that
\begin{equation}
M(u,r)=\frac{1}{6}r^3\Lambda^*(u).
\end{equation}
Then using this mass function in the rotating metric presented in
equation (6.4) of Ref. 2, we obtain a rotating metric, describing
a {\sl non-stationary} de Sitter model with cosmological term
$\Lambda^*(u)$ in the null coordinates ($u,r,\theta,\,\phi$) as
\begin{eqnarray}
ds^2&=&\Big\{1-{r^4\Lambda^*(u)\over3R^2} \Big\}\,du^2 +2du\,dr
\cr\cr &&+2a{r^4\Lambda^*(u)\over3R^2}\,{\rm
sin}^2\theta\,du\,d\phi -2a\,{\rm sin}^2\theta\,dr\,d\phi \cr\cr
&&-R^2d\theta^2 -\Big\{(r^2+a^2)^2-\Delta^*a^2\,{\rm
sin}^2\theta\Big\}\,R^{-2}{\rm sin}^2\theta\,d\phi^2,
\end{eqnarray}
where $R^2=r^2+a^2{\rm cos}^2\theta$ and
$\Delta^*=r^2-{r^4\Lambda^*(u)}/3+a^2$. Here $\Lambda^*(u)$
denotes an arbitrary non-increasing function of the retarded time
coordinate $u$ and $a$ being a constant rotational parameter. When
one sets the function $\Lambda^*(u)$ to be a constant, the line
element (3) will reduce to the {\sl rotating stationary} de Sitter
space-time [2]. The complex null vectors for the above metric can
be chosen as follows:
\begin{eqnarray}
&&\ell_a=\delta^1_a -a\,{\rm sin}^2\theta\,\delta^4_a, \cr\cr
&&n_a=\frac{\Delta^{*}}{2\,R^2}\,\delta^1_a+ \delta^2_a
-\frac{\Delta^{*}}{2\,R^2}\,a\,{\rm sin}^2\theta\,\delta^4_a,
\\\ &&m_a=-{1\over\surd 2R}\,\Big\{-ia\,{\rm
sin}\,\theta\,\delta^1_a+R^2\,\delta^3_a +i(r^2+a^2)\,{\rm
sin}\,\theta\,\delta^4_a\Big\}. \nonumber
\end{eqnarray}
Here $\ell_a$,\, $n_a$ are real null vectors and $m_a$ is complex
with the normalization conditions $\ell_an^a= 1 = -m_a\bar{m}^a$.
By virtue of Einstein's field equations, we calculate the
energy-momentum tensor describing matter field for the
non-stationary space-time as
\begin{eqnarray}
T_{ab} &=&\mu^*\,\ell_a\,\ell_b+ 2\,\rho^*\,\ell_{(a}\,n_{b)}
+2\,p\,m_{(a}\bar{m}_{b)}+ 2\,\omega\,\ell_{(a}\,\bar{m}_{b)} +
2\,\bar{\omega}\,\ell_{(a}\,m_{b)},
\end{eqnarray}
where
\begin{eqnarray}
&&\mu^* =-{r^4\over 6KR^2R^2}\Big\{2r\Lambda^*(u)_{,u} + a^2{\rm
sin}^2\theta\,\Lambda^*(u)_{,uu}\Big\}, \quad \rho^* = {r^4\over
KR^2R^2}\Lambda^*(u), \cr\cr &&p = -{r^2\Lambda^*(u)\over
KR^2R^2}\Big\{{r^2+2a^2\,{\rm cos}^2\theta}\Big\}, \quad \omega
=-{i\,a\,r^3{\rm sin}\,\theta\over6\surd
2KR^2R^2}\Big(R-3\bar{R}\Big)\Lambda^*(u)_{,u},
\end{eqnarray}
with the universal constant $K=8\pi G/c^4$. The quantity $\mu^*$
interprets as a null density based on the derivative of
$\Lambda^*(u)$; $\rho^*$ and $p$ are the density and pressure of
the non-stationary matter field, and $\omega$ represents the
rotational force density determined by the rotational parameter
$a$ coupling with the derivative of $\Lambda^*(u)$. That is, for
non-rotating model $a=0$,  the rotational density $\omega$ will
vanish, describing a non-rotating fluid model for non-stationary
de Sitter universe. From (6) we find the equation of state (the
ratio of the pressure to energy density ) for the non-stationary
rotating solution as $w = p/\rho=-(r^2+2a^2{\rm
cos^2\theta})r^{-2}$ for non-zero rotational parameter $a$. This
equation of state will take the value $w = -1$ at the poles
$\theta=\pi/2$ and $\theta=3\pi/2$, showing that the de Sitter
solution (3) with non-constant $\Lambda^*(u)$ describes a rotating
non-stationary dark energy model possessing negative pressure.

The trace of energy momentum tensor $T_{ab}$ (5) is found as
\begin{equation}
T=2(\rho^*-p).
\end{equation}
Here it is observed that $\rho^* - p>0$ for rotating
non-stationary de Sitter model. The energy-momentum tensor (5)
satisfies the energy conservation equations
$T^{ab}_{\;\;\;\;;b}=0$. The verification of these equation may be
seen in Appendix A below. The Ricci scalar $\Lambda$ $(\equiv
{1\over 24}g^{ab}\,R_{ab})$, describing matter field by virtue of
Einstein's field equations, is found as
\begin{equation}
\Lambda =\frac{1}{6}\,r^2\Lambda^*(u)R^{-2}.
\end{equation}
Other Ricci scalars are related with $\mu^*$, $\rho^*$, $p$ and
$\omega$ (6) as $K\mu^* = 2\,\phi_{22}$, $K\omega = -
2\,\phi_{12}$, $K\rho^* = 2\,\phi_{11} + 6\,\Lambda$ and $Kp =
2\,\phi_{11} - 6\,\Lambda$. The energy momentum tensor (5) also
satisfies all the three energy conditions: (i) weak,
$T_{ab}U^aU^b\geq0$, (ii) strong, $T_{ab}U^aU^b\geq {1\over2}T$
and (iii) dominant, for a time-like observer with its
four-velocity vector $U^a$ as shown in Ref. 2. It is noted that
$T_{ab}$ (5) does not describe a perfect fluid, i.e. for a
non-rotating perfect fluid, $T^{(\rm {pf})}_{ab} =
(\rho^*+p)u_a\,u_b-p\,g_{ab}$, with unit time-like vector $u_a$
and trace $T^{(\rm {pf})}=\rho^*-3p$, which is different from the
one given in (7).

The existing Weyl scalars, determining gravitational field for the
space-time metric (3) are obtained as
\begin{eqnarray}
&&\psi_2={r^2\Lambda^*(u)\over6\bar{R}\,\bar{R}
R^2}\{(r+2i\,a\,\cos \theta)\bar{R}-rR\}, \cr\cr
&&\psi_3=-{i\,a\,r^3{\rm sin}\theta\over{3\surd 2\bar{R}
\,\bar{R}\,R^2}}\,(r+\bar{R})\Lambda^*(u)_{,u}, \\\
&&\psi_4 ={{a^2\,r^4\,{\rm sin}^2\theta}\over 12\bar{R}
\,\bar{R}\,R^2R^2}\{R^2\Lambda^*(u)_{,uu}-
2\,r\,\Lambda^*(u)_{,u}\}. \nonumber
\end{eqnarray}
From the non-vanishing Weyl scalars $\psi_2$, $\psi_3$ and
$\psi_4$, it is observed that the rotating non-stationary de
Sitter model (3) is an algebraically special (precisely, type II:
$C_{abc[d}\ell_{h]}\ell^{b}\ell^{c}=0$, with $\psi_0=\psi_1=0$
[9]) in the Petrov classification of space-time with a null vector
$\ell_a$ (4), which is geodesic, shear free, expanding
($\hat{\theta}\equiv {1\over2}\ell^a_{\,;a}=rR^{-2}$) as well as
non-zero twist ($\hat{\omega}^2 \equiv
\frac{1}{2}\ell_{[a;\,b]}\ell^{a;\,b}
={a^2\cos^2\theta}R^{-2}R^{-2}$). Here the function $\Lambda^*(u)$
does not involve in the expression of expansion $\hat{\theta}$ and
twist $\hat{\omega}^2$ of the null vector $\ell^a$. This means
that the physical properties of this null vector $\ell^a$ are same
for both {\sl stationary} [2] as well as {\sl non-stationary} (3)
rotating de Sitter models, though they have different
gravitational fields with different energy momentum tensors.

The expressions of $\psi_3$ and $\psi_4$ above involve the
derivative of $\Lambda^*(u)$ as $\Lambda^*(u)_{,u}$, coupling with
the rotational parameter $a$. So at some point when $\Lambda^*(u)$
sets to be a constant for non-zero rotation ($a\neq 0$), both
$\psi_3$ and $\psi_4$ will vanish. At that point the gravitational
field of the observer will be of type $D$ ($\psi_2\neq 0$) in the
Petrov classification of stationary space-time. That is, the
space-time becomes the rotating stationary de Sitter solution with
cosmological constant $\Lambda^*$ [2].

The metric (3) can be expressed in the coordinate system
 $(t,\,x,\,y,\,z)$ as
\begin{eqnarray*}
ds^2&=&dt^2-dx^2-dy^2-dz^2 \cr\cr
      &&+{\{r^6\Lambda^*(t,r)\} \over 3(r^4+a^2\,z^2)}\,\Big[dt-
      {1\over(r^2+a^2)}\,\{r(xdx+ydy)+a(xdy-ydx)\}-{1\over
      r}\,zdz\Big]^2,
\end{eqnarray*}
where $r$ is defined, in terms of $x,\,y$ and $z$ [9]
\begin{eqnarray*}
r^4-(x^2+y^2+z^2-a^2)\,r^2-a^2\,z^2=0,\,
\end{eqnarray*}
with the following transformations
\begin{eqnarray*} &&x=(r\,{\rm
cos}\phi + a\,{\rm sin}\phi)\,{\rm sin}\theta,\cr &&y=(r\,{\rm
sin}\phi - a\,{\rm cos}\phi)\,{\rm sin}\theta, \cr &&z=r\,{\rm
cos}\theta,\;\;\; t=u+r.
\end{eqnarray*}
Then, the above transformed metric is in the Kerr-Schild form with
\begin{equation}
g^{\rm dS}_{ab} = \eta_{ab} + 2\,H(t,\,x,\,y,\,z)\,\ell_a \ell_b,
\end{equation}
where $\eta_{ab}$ is the flat metric and
\begin{eqnarray*}
&&H(t,\,x,\,y,\,z)=-{r^6\Lambda^*(t,r)\over 3(r^4+a^2\,z^2)}, \cr
&&\ell_a\,dx^a=dt-
      {1\over(r^2+a^2)}\,\{r(xdx+ydy)+a(xdy-ydx)\}-{1\over r}\,zdz.
\end{eqnarray*}
In $u$-coordinate system the null vector $\ell_a$ is given in (4)
above. The Kerr-Schild ansatz (10) confirms that the rotating
non-stationary de Sitter model (3) is a solution of Einstein's
field equations of non-constant curvature.

The rotating non-stationary de Sitter solution has an apparent
singularity when $\Delta^*=r^2-{r^4\Lambda^*(u)}/3+a^2=0$. This
equation has four roots $r_{+\,+}$, $r_{+\,-}$, $r_{-\,+}$ and
$r_{-\,-}$. They are found as
\begin{eqnarray}
r_{\pm(\pm)}=\pm\sqrt{\frac{1}{2\Lambda^*(u)}\Big\{3\pm\sqrt{9
+12a^2\Lambda^*(u)}\,\Big\}}.
\end{eqnarray}
Now let us denote  these roots $r_{+\,+}$, $r_{+\,-}$, $r_{-\,+}$,
$r_{-\,-}$ as $r_1$, $r_2$, $r_3$, $r_4$ respectively (for the
simplicity, $r_j, \;\, j=1,2,3,4$). Then, these roots have the
following relation
\begin{eqnarray*}
(r-r_1)(r-r_2)(r-r_3)(r-r_4)=
-\,\frac{3}{\Lambda^*(u)}\Big\{r^2-\frac{r^4\Lambda^*(u)}{3}+a^2\Big\}.
\end{eqnarray*}
Then each root represents the location of the cosmological horizon
for the observer and associates  an area of the horizon at each
point at $r=r_j, \;\, j=1,2,3,4$,
\begin{eqnarray}
{\cal A}_j &=&
\int_{0}^{\pi}\int_{0}^{2\pi}\sqrt{g_{\theta\theta}g_{\phi\phi}}
\,d\theta\,d\phi\,\Big|_{r=r_{j}} \cr\cr &=& 4\pi\{r_{j}^2 +a^2\}.
\end{eqnarray}
According to Bekenstein-Hawking area-entropy formula [3], these
areas ${\cal A}_j$ will determine the entropies ${\cal S}_j$ of
the horizons of the de Sitter model (3) by the relation ${\cal
S}_j={\cal A}_j/4$ [3]. Thus, we find them as
\begin{eqnarray}
{\cal S}_j=\pi\{r_{j}^2 +a^2\}.
\end{eqnarray}
The gravity of the cosmological horizons is determined by the
surface gravity, defined by $\kappa n^a= n^b\,\nabla_{b}\,n^a$ in
[5], where the null vector $n^a$ given in (4) above is
parameterized by the coordinate $u$, such that
$d/du=n^a\nabla_{a}$, and has the normalization condition
$\ell_an^a=1$ with the null vector $\ell_a$. The surface gravities
$\kappa_j$ associated at each $r=r_{j}$ are found as  below:
\begin{eqnarray}
\kappa_{p}&=&\frac{r_p}{3R^2_p}\Lambda^*(u)\{r_p+r_{p+1}\}\{r_{p+1}-r_p\}
\;\; {\rm for}\;\; p=1,3, \\
\kappa_{p}&=&\frac{r_p}{3R^2_p}\Lambda^*(u)\{r_{p-1}+r_p\}\{r_{p-1}-r_p\}
\;\; {\rm for}\;\; p=2,4,
\end{eqnarray}
where $R^2_j=r^2_j+a^2{\rm cos}^2\theta,\;\, j=1,2,3,4$. The
surface gravities $\kappa_j$ may be regarded as the gravitational
field on the cosmological horizons. From (14) and (15), we observe
that $\kappa_1$ and $\kappa_2$ will be zero when $r_1$ and $r_2$
coincide. Similarly,  $\kappa_3$ and $\kappa_4$ will vanish when
$r_3=r_4$.

The coincidence of two roots $r_1$ and $r_2$ leads to a condition
that $\{9+12a^2\Lambda^*(u)\}=0$. This condition implies that
$r_3$ and $r_4$ also coincide. Then all roots take the form $r_1 =
r_2 = -r_3 = -r_4 = \surd\{3/(2\Lambda^*(u))\}$. Accordingly, the
area of the horizon at each point $r_1$, $r_2$, $r_3$ and $r_4$
are found as
\begin{eqnarray}
{\cal A}_j = \frac{3\pi}{\Lambda^*(u)}.
\end{eqnarray}
This implies that the entropy associated with each point becomes
\begin{eqnarray}
{\cal S}_j=\frac{3\pi}{4\Lambda^*(u)},\qquad j = 1, 2, 3, 4.
\end{eqnarray}
From this expression we find that the entropies ${\cal S}_j$ at
$r_j$ are inversely proportional to the cosmological function
$\Lambda^*(u)$. It is to mention that the value of $\Lambda^*(u)$
is supposed to reduce according to the retarded time $u$ change.
Consequently the entropies ${\cal S}_j$ may increase, as the
function $\Lambda^*(u)$ reduces. It is found that once the
function $\Lambda^*(u)$ takes the constant value, as in the case
of {\it stationary} rotating de Sitter universe [2], the entropies
${\cal S}_j$ associated with  $r_j$ will take constant values.

The angular velocities $\Omega_j$ for the horizons are found as
\begin{eqnarray}
\Omega_j &=& \lim_{r \to
r_j}\Big(-\frac{g_{u\phi}}{g_{\phi\phi}}\Big)
=-\frac{a\,r^4_j\,\Lambda^*(u)}{3(r^2_j+a^2)^2}.
\end{eqnarray}
The coincidences of the roots ($r_1 = r_2$, and $r_3 = r_4$) imply
that $r_j= 3/(2\Lambda^*(u))$. Then the angular velocities take
the forms
\begin{eqnarray}
\Omega_j = -\frac{4a}{3}\Lambda^*(u).
\end{eqnarray}
This indicates that angular velocities are directly proportional
the cosmological function $\Lambda^*(u)$, and the effect of the
change in $\Lambda^*(u)$ will certainly affect on the angular
velocities associated with $r_j$. It is also emphasized that the
angular velocity $\Omega_j$ given in (18) will vanish when the
rotational parameter $a$ tends to zero, showing the fact that
there is no angular velocity for non-rotating de-Sitter space-time
with vanishing rotational parameter $a=0$.

It is quite interesting to discuss the nature of the non-rotating
($a=0$) non-stationary de Sitter model with $\Lambda^*(u)$.
Although the non-rotating  de Sitter model does not explain the
complete structure of the space-time, the metric is very simple
without much mathematical expressions. Thus, when one sets the
rotational parameter $a$ to zero, the metric (3) reduces to
non-rotating de Sitter model as
\begin{eqnarray}
ds^2&=&\Big\{1-\frac{1}{3}r^2\Lambda^*(u)\Big\}\,du^2 +2du\,dr
-r^2(d\theta^2+{\rm sin}^2\theta\,d\phi^2).
\end{eqnarray}
In this situation, the Weyl scalars $\psi_2$, $\psi_3$ and
$\psi_4$ given in (9) are vanished showing that the space-time
becomes the conformally flat ($C_{abcd}=0$). Then the
energy-momentum tensor (5) takes the form
\begin{equation}
KT_{ab}=-\frac{1}{3}r\Lambda^*(u)_{,u}\ell_a\ell_b +
\Lambda^*(u)g_{ab}
\end{equation}
with its trace $KT=4\Lambda^*(u)$. Here the energy-momentum tensor
(21) involves a Vaidya-like null radiation term
$-\frac{1}{3}r\Lambda^*(u)_{,u}\ell_a\ell_b$, which will vanish
when $r\rightarrow 0$, and satisfies the energy conservation
equation $T^{ab}_{\;\;\;\,;b}=0$. However, it still maintains the
non-stationary behavior $\Lambda^*(u)\neq$ constant, showing that
the space-time of the observer is naturally time dependent even at
$r\rightarrow 0$. The energy momentum tensor (21) will become the
one of the original de Sitter model [1] when $\Lambda^*(u)$ takes
a constant value. Using the energy momentum tenor (21) we find
Einstein's field equations $G_{ab}=-KT_{ab}$ for the non-rotating
metric (20) as follows
\begin{eqnarray}
R_{ab} - \frac{1}{2}\,R\,g_{ab}  + \Lambda(u)g_{ab} =
-T_{ab}^{(\rm NS)},
\end{eqnarray}
where the non-stationary evolution part of the energy-momentum
tensor (20) is given by
\begin{eqnarray*}
T_{ab}^{(\rm NS)}=-\frac{1}{3}r\Lambda(u)_{,u}\ell_a\ell_b,
\end{eqnarray*}
which has zero-trace $T^{(\rm NS)}=0$. It is to emphasize that the
universal constant $K$ does not involve in the field equations
(22). For future use we also present the null energy density
$\mu^*$, energy density $\rho^*$ and pressure $p$ of the
non-rotating de Sitter metric as follows
\begin{eqnarray} \mu^* =-\frac{r}{K} \Lambda^*(u)_{,u} \quad
\rho^* = \frac{\Lambda^*(u)}{K}, \quad p =
-\frac{\Lambda^*(u)}{K}.
\end{eqnarray}
From (23) we find the equation of state $w=p/\rho^*=-1$ with the
negative pressure of variable $\Lambda^*(u)$. This shows the fact
that our non-stationary de Sitter solution (20) is in agreement
with the cosmological constant ($\Lambda^*$) de Sitter solution
possessing the equation of state $w=-1$ in the dark energy
scenario [10, 11, 12].

The metric (20) has an apparent singularity with a horizon at
$r_{\pm}=\pm\{3\Lambda^{*(-1)}(u)\}^{1/2}$. The entropies ${\cal
S}_{\pm}$ and the surface gravities $\kappa_{\pm}$ of the horizon
associated with $r=r_{\pm}$ for the non-rotating metric (20) are
found as
\begin{eqnarray}
{\cal S}_{\pm}=\frac{3\pi}{\Lambda^*(u)} \quad {\rm and} \quad
\kappa_{\pm}= \pm\sqrt\frac{\Lambda^*(u)}{3}.
\end{eqnarray}
From these expressions we observe that ${\cal S}_{\pm}$ are
inversely proportional to $\Lambda^*(u)$, whereas $\kappa_{\pm}$
are not. It is emphasized that, according to the change of the
retarded time $u$, the value of $\Lambda^*(u)$ may decrease. This
ensures that the entropies ${\cal S}_{\pm}$ of the non-stationary
de Sitter model will increase, whereas the surface gravities
$\kappa_{\pm}$ decrease. Such observations of changes of the
values of ${\cal S}_{\pm}$ and $\kappa_{\pm}$ can be found only in
the case of non-stationary de Sitter model (20). This is the
important aim of the study of non-stationary de Sitter space-time.
Our results of the non-rotating de Sitter  solution are in
agreement with those of Gibbons and Hawking [3], when the
cosmological function $\Lambda^*(u)$ tends to a constant
$\Lambda^*$.

The Kretschmann scalar for non-rotating de Sitter model (17) takes
the form
\begin{equation}
{\cal K} \equiv R_{abcd}R^{abcd}=\frac {8}{3}\Lambda^*(u)^2,
\end{equation}
which does not involves any derivative term of $\Lambda^*(u)$, and
will not change its value at $r\rightarrow 0$ and $r\rightarrow
\infty$. The above Kretschmann scalar will become the one of
original de Sitter model when $\Lambda^*(u)$ takes a constant
value. That is, though the energy momentum tensor  (21) for
non-rotating non-stationary model is found different from the one
of non-rotating stationary de Sitter solution, the forms of
Kretschmann scalar for both {\sl non-rotating non-stationary} (20)
and {\sl stationary} models [1] have similar structures with a
difference in nature of the cosmological function $\Lambda^*(u)$.

Our result discussed here includes the following cosmological
models: (i) original de Sitter when $a=0$, $\Lambda^*(u)$=
constant [1], (ii) rotating stationary de Sitter when $a\neq 0$,
$\Lambda^*(u)$= constant [2], (iii) non-rotating non-stationary de
Sitter when $a= 0$, $\Lambda^*(u)\neq$ constant, (iv) rotating
non-stationary de Sitter when $a\neq 0$, $\Lambda^*(u)\neq$
constant. From the study of non-stationary models it is observed
that the gravitational field of the space-time of the rotating
model (3) is algebraically special in the Petrov classification
whereas the non-rotating one (20) is conformally flat with the
energy momentum tensor describing non-empty space. It is hoped
that known stationary works, rotating or non-rotating, with
cosmological constant, may be extended to the non-stationary ones
by using the non-stationary de Sitter model, rotating (3) or
non-rotating (20). It is to note that to the best of the author's
knowledge, these non-stationary de Sitter cosmological solutions
(3) and (20) have not been seen discussed before.

\section*{Acknowledgments}

The author is thankful the Inter University Centre for Astronomy
and Astrophysics (IUCAA), Pune for hospitalities during his visit.
This work is supported by the University Grants Commission (UGC),
New Delhi, File No. 31-87/2005 (SR). The comment of a referee of
this article is highly appreciable.

\begin{appendix}
\setcounter{equation}{0}
\renewcommand{\theequation}{A\arabic{equation}}

\section*{Appendix A: Energy-momentum tensor conservation equations}

In this appendix we establish the fact that the energy-momentum
tensor (5) satisfies the conservation equations
$T^{ab}_{\;\;\;\;;b}=0$. These are four equations, which can
equivalently be expressed in three equations -- two are real and
one complex, using Newman-Penrose spin coefficients. Hence, we
find the following
\begin{eqnarray}
D\rho^{*}=(\rho^{*}+ p)(\rho+\bar{\rho})+\omega \bar{\kappa}^*+
\bar{\omega} \kappa^*,
\end{eqnarray}
\begin{eqnarray}
D\mu^*+\nabla \rho^* + \bar{\delta}\omega + \delta \bar{\omega}
&=& \mu^*\{(\rho+\bar{\rho})- 2 (\epsilon+\bar{\epsilon})\}
-(\rho^{*}+ p)(\mu+\bar{\mu})\cr &&- \omega(2\pi+ 2\bar{\beta} -
\bar{\tau}) -\omega(2\bar{\pi}+ 2\beta- \tau),
\end{eqnarray}
\begin{eqnarray}
D\omega+\delta p=\mu^*\kappa^*+(\rho^*+p)(\tau-\bar{\pi})
+\bar{\omega}\sigma-\omega(2\bar{\epsilon}-2\rho-\bar{\rho})
\end{eqnarray}
where $\kappa^*$, $\tau$, $\pi$, etc. are spin coefficients, and
the derivative operators are defined by
\begin{eqnarray}
D \equiv \ell^a \partial_a, \quad \nabla \equiv n^a \partial_a,
\quad\delta \equiv m^a \partial_a, \quad \bar{\delta}\equiv
\bar{m}^a \partial_a.
\end{eqnarray}
These are general equations for an energy-momentum tensor of the
type (5).

Now, in order to verify whether components of $T^{ab}$ with the
quantities $\mu^*$, $\rho^*$, $p$ and $\omega$ given in (6) for
the rotating non-stationary de Sitter metric satisfy these
conservation equations (A1--A3) or not, we present the NP spin
coefficients for the metric (3):
\begin{eqnarray}
&&\kappa^*=\sigma=\lambda=\epsilon=0, \cr\cr
&&\rho=-\frac{1}{\bar{R}}, \quad
\mu=-\frac{\Delta^*}{2\bar{R}\,R^2},\cr\cr
&&\alpha=\frac{(2ai-R\,\cos\theta)}{2\surd 2\bar{R}
\,\bar{R}\,\sin\theta},\quad \beta=\frac{\cot\theta}{2\surd 2\,R},
\cr\cr &&\pi={i\,a\,{\rm sin}\,\theta\over{\surd
2\bar{R}\,\bar{R}}}, \quad
\tau=-{i\,a\,{\rm sin}\,\theta\over{\surd 2R^2}},\\\
&&\gamma=\frac{1}{2\,\bar{R}\,R^2}\, \Big [\{r -
\frac{2}{3}r^2\Lambda^*(u)\}\bar{R} - \Delta^*\Big ], \cr\cr
&&\nu=\frac{1}{6\surd
2\,\bar{R}\,R^2}\,iar^4\sin\theta\Lambda^*(u)_{,u}  \nonumber
\end{eqnarray}
The derivative operators (A4) are given as follows:
\begin{eqnarray}
&&D=\partial_r, \cr &&\nabla=\frac{1}{R^2}\Big\{(r^2 +
a^2)\,\partial_u
-\frac{\triangle^*}{2}\partial_r + a\,\partial_{\phi}\Big\}, \\\
&&\delta=\frac{1}{\surd 2\,R}\Big\{i\,a\,\sin\theta\,\partial_u
+\partial_{\theta} +
\frac{i}{\sin\theta}\,\partial_\phi\Big\},\nonumber
\end{eqnarray}
where $\Delta^*=r^2-{r^4\Lambda^*(u)}/3+a^2$. The equations (A1)
and (A3) are comparatively easier to verify than (A2). Therefore,
we shall not show their verification here except for the equation
(A2). Now, by virtue of (6) and (A6), the left side of (A2) is
found as
\begin{eqnarray}
D\mu^*+\nabla \rho^* + \bar{\delta}\omega + \delta \bar{\omega}
&=&\frac{r^5a^2\sin^2\theta}{3KR^2R^2R^2}\,\Lambda^*(u)_{,uu} -
\frac{2r^3a^2\cos^2\theta\triangle^*}{KR^2R^2R^2R^2}\,\Lambda^*(u)
\cr\cr &&+ \Big\{\frac{2r^6}{3KR^2R^2R^2} -
\frac{a^2r^4}{3KR^2R^2R^2} \cr\cr && + \;
\frac{2a^2r^4\sin^2\theta}{3KR^2R^2R^2R^2}\Big(3a^2\cos^2\theta
+r^2\Big)\Big\}\Lambda^*(u)_{,u}.
\end{eqnarray}
which can be shown equal to the right side of (A2), by using (6)
and (5A). This leads to the conclusion of the verification that
the energy-momentum tensor (5) satisfies the condition
$T^{ab}_{\;\;\;\,;b}=0$.
\end{appendix}

\end{document}